\documentclass[aps,pra,twocolumn,showkeys]{revtex4-1}

\usepackage{graphicx}
\usepackage{amsthm,amssymb,amsmath}
\usepackage{textcomp}
\usepackage{xspace}
\usepackage{color}

\begin{document}


\title{Quantum teleportation and nonlocality:\\the puzzling predictions of entanglement are coming of age}

\author{Nicolas Gisin}%
\email{nicolas.gisin@unige.ch}
\affiliation{Group of Applied Physics, University of Geneva, CH-1211 Geneva 4, Switzerland
}

\author{S\'ebastien Tanzilli}%
\email{sebastien.tanzilli@unice.fr}
\affiliation{Universit\'e Nice Sophia Antipolis, Laboratoire de Physique de la Mati\`ere Condens\'ee, CNRS UMR 7336, Parc Valrose, 06108 Nice Cedex 2, France.
}

\author{Wolfgang Tittel}%
\email{wtittel@ucalgary.ca}
\affiliation{Institute for Quantum Science and Technology, and Department of Physics and Astronomy,\\University of Calgary, Canada
}

\date{\today}


\keywords{Entanglement, Nonlocality, Quantum Teleportation\\
\,\\
\textit{Note: This paper is intended to be published in the 2015 fourth issue of Europhysics News as a ``Feature Article''.}}

\maketitle

\section{Foreword}

Entanglement is the physical property that marks the most striking deviation of the quantum from the classical world. It has been mentioned first by the great Austrian Physicist Erwin Sch\"odinger in 1935 (an introduction to this and other quantum phenomena is [1]). Yet, despite this theoretical prediction now being 80 years old, and the famous experimental verifications by Alain Aspect dating back to the early eighties [2], entanglement and its use entered mainstream physics as a key element of quantum information science [3] only in the 1990's.

\section{Introduction}

The academic research into entanglement nicely illustrates the interplay between fundamental science and applications, and the need to foster both aspects to advance either one. For instance, the possibility to distribute entangled photons over tens or even hundreds of kilometers is fascinating because it confirms the quantum predictions over large distances, while quantum theory is often presented to apply to the very small (see Figure 1). On the other hand, entanglement enables quantum key distribution (QKD) [1]. This most advanced application of quantum information processing allows one to distribute cryptographic keys in a provably secure manner. For this, one merely has to measure the two halves of an entangled pair of photons. Surprisingly, and being of both fundamental and practical interest, the use of entanglement removes even the necessity for trusting most equipment used for the measurements [5]. Furthermore, entanglement serves as a resource for quantum teleportation (see Figure 2) [1]. In turn, this provides a tool for extending quantum key distribution to arbitrarily large distances and building large-scale networks that connect future quantum computers and atomic clocks [6].

\begin{figure}
\includegraphics[width=\columnwidth]{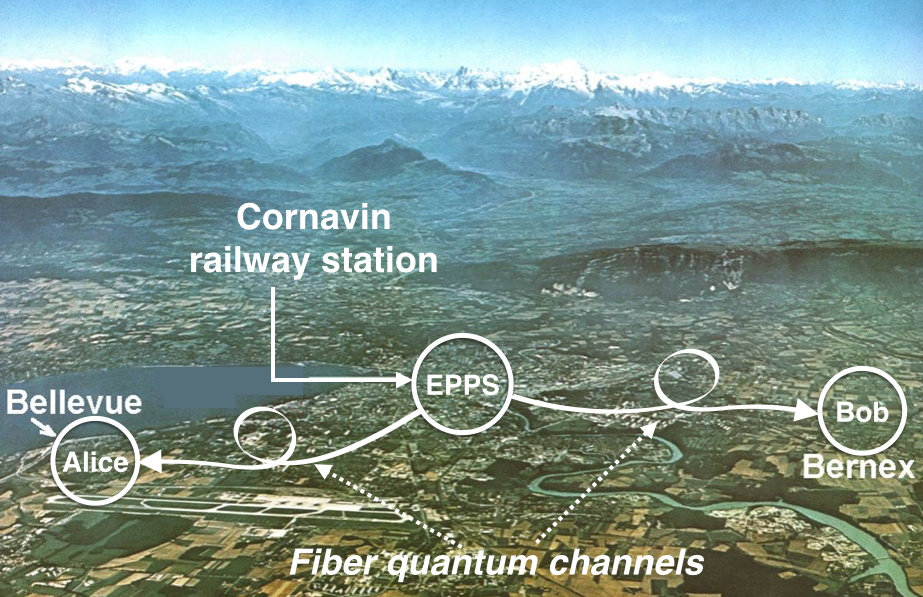}
\caption{Long-distance entanglement distribution. A source, located at the Geneva main railway station, distributes entangled photon pairs emitted at a telecom wavelength to two remote villages (Alice's and Bob's station located in Bellevue and Bernex, respectively, in the Geneva back-country). Those two user stations are more than 10\,km away from each other, and are connected to the source through commercial fiber optics quantum channels. This experiment reported the first tests of entanglement distribution and nonlocality in the real world [4]. Entanglement can be further exploited for establishing secret sequences of bits (\textit{i.e.}, secret keys) finding applications in cryptography. EPPS: entangled photon pair source.}
\end{figure}

In the following, we describe the counter-intuitive properties of entangled particles as well as a few recent experiments that address fundamental and applied aspects of quantum teleportation. While a lot of work is being done using different quantum systems, including trapped ions, color centers in diamond, quantum dots, and superconducting circuits, we will restrict ourselves to experiments involving photons due to their suitability for building future quantum networks.

\section{Entanglement -- a puzzling consequence of quantum theory}

Entanglement manifests itself through highly, if not perfectly, correlated results of measurements performed on particles that can in principle be arbitrarily far away - say, one on the earth, and one on the moon. If we consider two photons and their polarization as an example, quantum theory describes, and experiments confirm, that photon one might be found horizontally polarized, and photon two vertically polarized. Or vice versa. What is more, one might also find that one photon is polarized at 45\textdegree and the other one at -45\textdegree. All that is known in advance is that the two photons are orthogonally polarized.

What is particularly discomforting about these correlations is that they cannot be explained by attributing properties to the individual photons that determine,
irrespectively of what happens to the other photon, how each will respond to its measurement. Entangled particles behave in unison, regardless of their separation and even if they are measured simultaneously. The result of a quantum measurement is random, but, somehow, this random event manifests itself at both locations. Einstein called this ``spooky action at a distance'', though there is no real action from one side onto the other [7]! But while the question of how to understand this invisible, nonlocal tie remains intriguing, the tie enables applications of quantum communication such as teleportation.

\section{Quantum teleportation -- a surprising possibility}

Suppose you see a beautiful sculpture in a museum and you would like to have the same at home. Unfortunately, you can't take it with you. However, you can accurately measure all its properties - \textit{e.g.}, its shape (height, length and depth) - and then reproduce an identical copy for your living room. But this ``measure-and-reproduce'' strategy would fail if the sculpture was a photon. Indeed, for the case of a quantum object, the quantum no-cloning theorem [1] tells us that perfect copying of, say, the photon's polarization is impossible.

Quantum teleportation (see [8] for a recent review article), proposed in 1993 and first experimentally demonstrated at the Universities of Rome and Innsbruck in 1997, allows the flawless transfer of a property between two quantum particles - \textit{e.g.}, two photons - without running into a contradiction with the no-cloning theorem. As explained in Figure 2a, it requires three particles - one whose property is to be teleported (A), and two that are originally entangled (C and B). The comparative measurement of the property of the first photon with that of one member of the entangled pair allows transferring it from the first photon onto the second member of the pair. This measurement - the so-called Bell-state measurement - is named after the Northern Irish Physicist John Bell, who has played a seminal role in establishing the science of entanglement. The first photon loses its property during this measurement, that is, there is no 'copying' during teleportation. Moreover, the transfer does not happen instantaneously (something that is often claimed erroneously in the non-scientific literature), and hence there is no contradiction with another pillar of modern physics either - that of special relativity.

\begin{figure}
\centering
\begin{tabular}{c}
\includegraphics[width=\columnwidth]{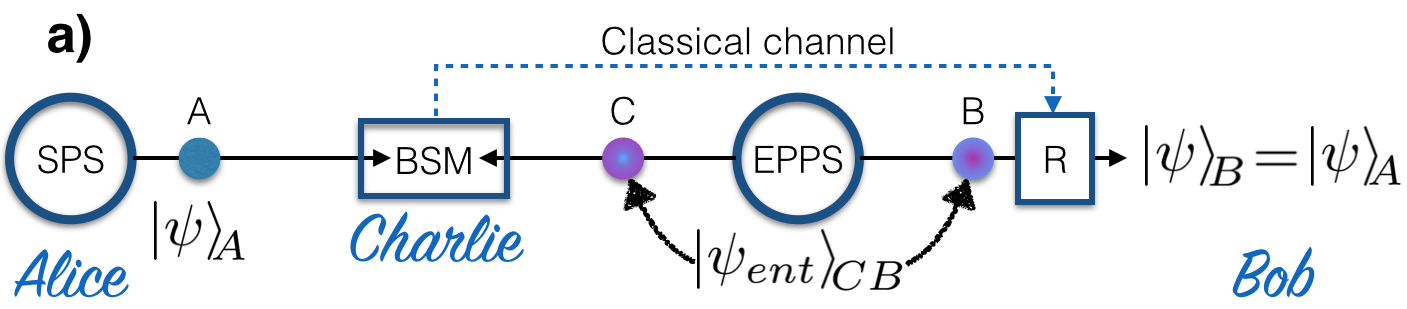}\\
\,\\
\includegraphics[width=\columnwidth]{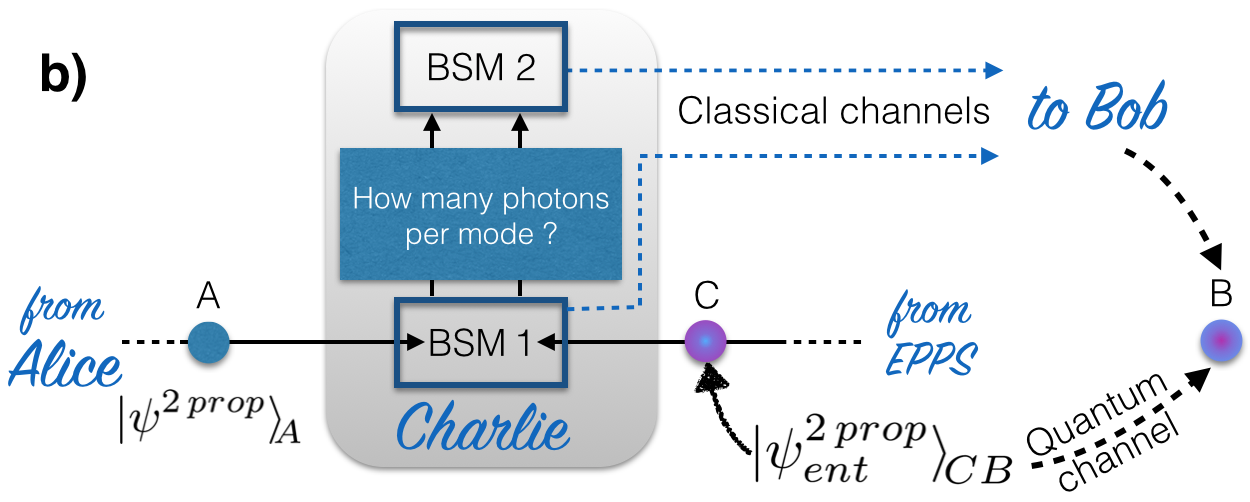}
\end{tabular}
\caption{Quantum teleportation. a) The original scheme. Photon A emitted on Alice's side, whose polarization, represented by $|\Psi\rangle_A$, should be teleported onto photon B at Bob's, is measured jointly with photon C. This joint measurement, called Bell-state measurement (BSM) reveals, loosely speaking, the difference in the electric field directions, without telling the individual directions. For instance, if we find that A and C are orthogonally polarized, and knowing from the original entanglement that the polarization of photon C is orthogonal to that of photon B, we find that the electric field of photon B must point into the same direction as that of photon A before the measurement. Note that the outcome of the BSM could also have been different, for example, A and C are identically polarized. Similar reasoning would then lead to the conclusion that photon B's polarization is orthogonal to that of photon A. Therefore, rotating it back would also allow one to perfectly recover the original polarization encoded into photon A. In short, the BSM, possibly followed by a well-defined rotation of the (unknown) polarization of photon B, allows one to teleport without error the property ``polarization'' from photon A onto photon B.
EPPS: entangled photon pair source; SPS: single photon source; R: polarization rotator. b) Teleportation of 2 properties. This scenario is comparable to that described in a) with, however, the possibility of teleporting 2 quantum properties coded on the same original single photon. In this case, the single BSM is replaced by two cascaded joint measurements, one for each property, but the second one is conditioned on the success of the first one.}
\end{figure}

\section{An application of quantum teleportation -- extending the reach}

Only five years after its discovery, it was realized that quantum teleportation is not only an intriguing manifestation of the puzzling predictions of quantum theory. Together with the possibility to transfer properties from flying photons onto stationary particles (\textit{i.e.}, to create quantum memory for light), it allows establishing entanglement over theoretically arbitrarily long distances by means of quantum repeaters [8]. This would allow building ultra-long-distance QKD links as well as quantum networks across countries, continents, or even the globe.

The first step in this direction was demonstrated in 2003 at the University of Geneva, when the photon that received the teleported property (photon B in Figure 2a) was sent over 2\,km of spooled, standard telecommunication fiber before being
measured [9]. This demonstration was extended in 2007 to more than 800\,m of deployed fiber that was part of the Swisscom fiber network [10] (see Figure 3a). In this experiment the receiving photon was already hundreds of meters away when the qubit to be teleported was prepared. More recently, the transmission distance of photon B was extended to more than 100\,km of air by researchers at the University of Vienna as well as the University of Science and Technology in Hefei [8].

\begin{figure}
\centering
\begin{tabular}{c}
\includegraphics[width=\columnwidth]{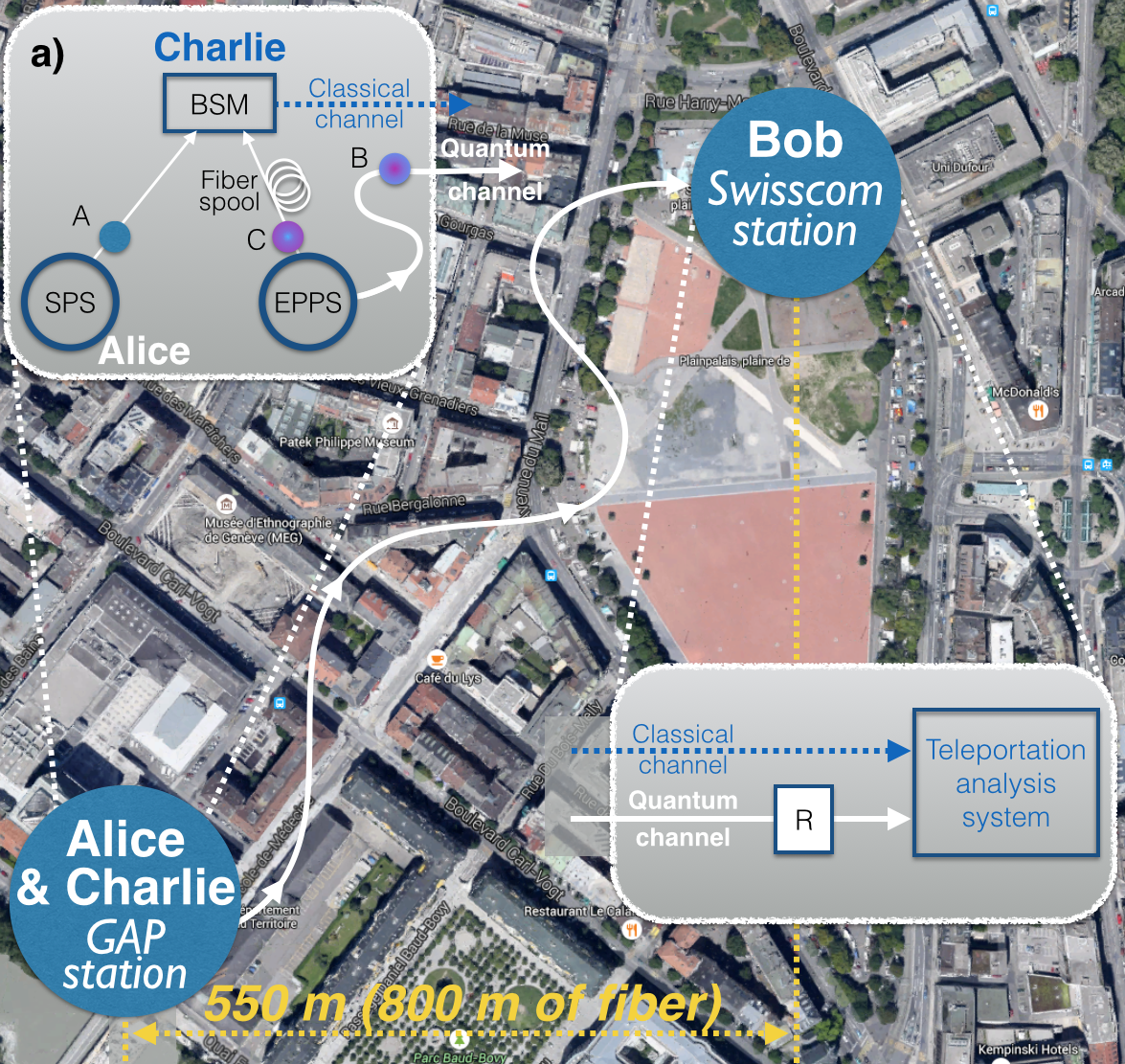}\\
\,\\
\includegraphics[width=\columnwidth]{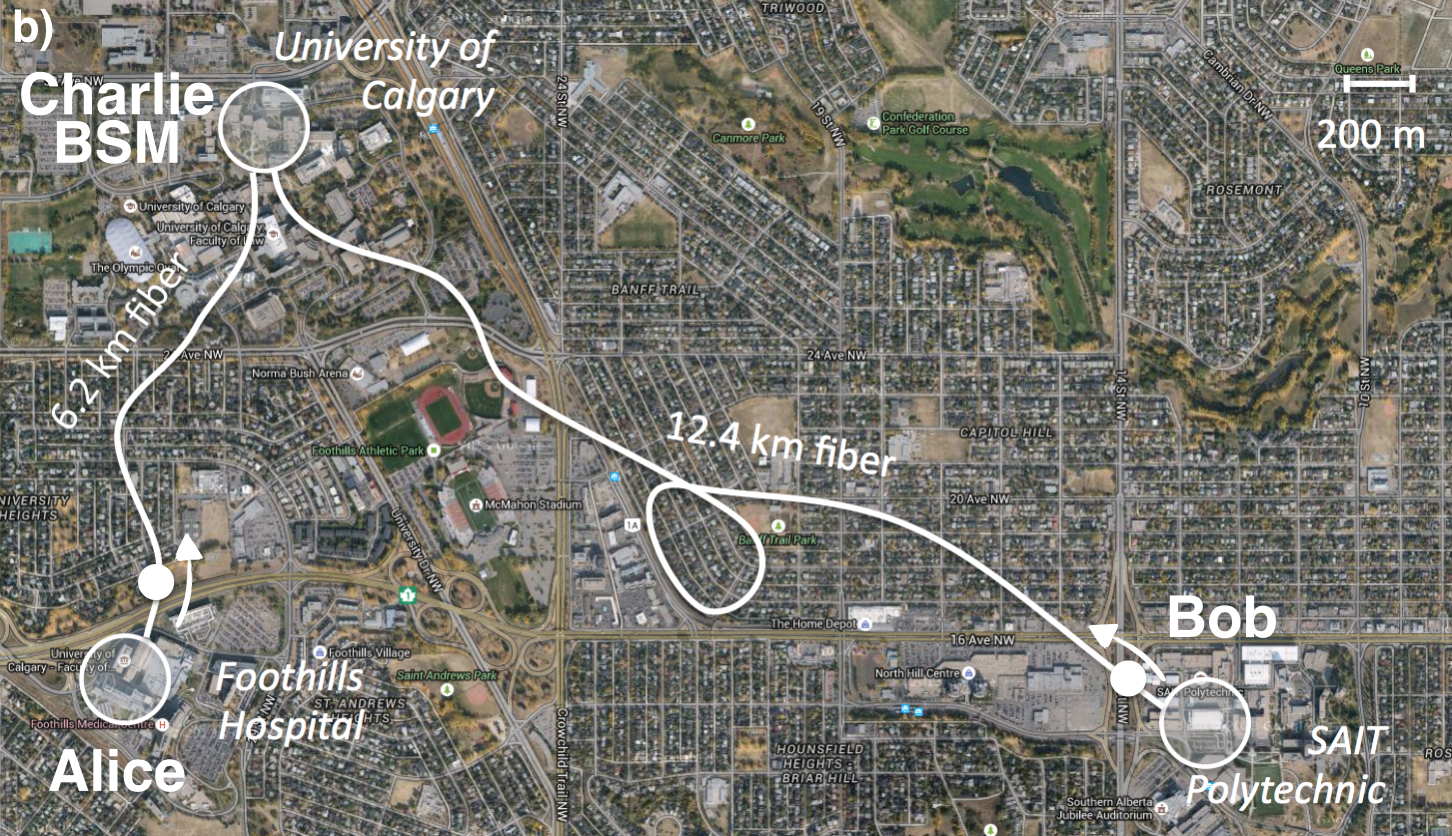}
\end{tabular}
\caption{Extending the reach. a) The 2007 Geneva teleportation experiment as an example in which photon B is already outside the lab when photon A is created, and continues to travel before being measured at Bob's [10]. b) The Calgary BSM as an example in which photons A and C are travelling a long distance before being jointly measured.}
\end{figure}

The next step is to extend the distance over which the other two photons (photon A and C in Figure 2a) travel before meeting for the Bell-state measurement. However, due to the difficulty of avoiding any modification of either photon during transmission, this challenge has not yet been met outside a well-controlled laboratory. An important step in this direction has been the recent demonstration by researchers at the University of Calgary of a Bell-state measurement - not a full quantum teleportation experiment - with photons that have been created at different places within the city of Calgary and travelled through the standard telecommunication fiber network before being measured [11] (see Figure 3b).

Another key achievement on the path towards a quantum network has been last year's teleportation at the University of Geneva of a property from a photon into a rare-earth-ion doped crystal, which stored it for 50 nanoseconds [12]. Notably, both photons that took part in the Bell-state measurement travelled over 12.5\,km of spooled fiber, thereby also meeting the above-described requirement for building extended quantum networks. This demonstration built on the previous observation that such crystals are suitable for storing members from entangled photon pairs, which was reported in 2011 by researchers in Geneva as well as in Calgary [13].

\section{A fundamental question -- teleportation of multiple properties}

With very few exceptions, all quantum teleportation experiments to date demonstrated the transfer of a single property of one particle, \textit{e.g.}, the polarization of a photon. However, objects encountered in our every-day life are composed of many elementary building blocks, \textit{e.g.}, many atoms, each of which being described by several properties. A natural question is therefore if one can teleport more complex quantum systems as well. Indeed, this is possible. A first guess may lead to the idea of teleporting all properties individually in a straightforward generalization of the scheme shown in Figure 2a. This guess is correct - most surprisingly even if the properties are entangled!

However, in the case of teleporting several properties encoded into the same particle, there is an interesting twist - at least in the scheme employed by researchers from the University of Science and Technology in Hefei in 2015 to transfer the angular orbital momentum and the polarization of a single photon [14]. As shown in Figure 2b, it requires, as an intermediate step, the confirmation that exactly one photon traveled along each of the two paths connecting the measurement that teleports the first property (the orbital angular momentum) with the measurement that teleports the second one (the polarization). Obviously, the use of standard single photon detectors is not a viable solution, as the photons would be destroyed during the measurements. But interestingly a standard 'single property teleporter' does the job. Indeed, a successful Bell-state measurement does not only transfer a property from one photon onto another photon (a process that is not distinguishable from the direct transmission of the original photon through the teleporter), but also confirms that the photon existed! This idea generalizes from two to any number of properties: an n-property teleporter requires an (n-1)-property teleporter as a plug-in, which itself requires an (n-2)-property teleporter, \textit{etc.} Figure 4 shows a fully integrated 'on-chip' realization of a one-property teleporter, developed at the University of Nice [15], that would constitute a good starting point for building such a nested scheme.

\begin{figure}
\includegraphics[width=\columnwidth]{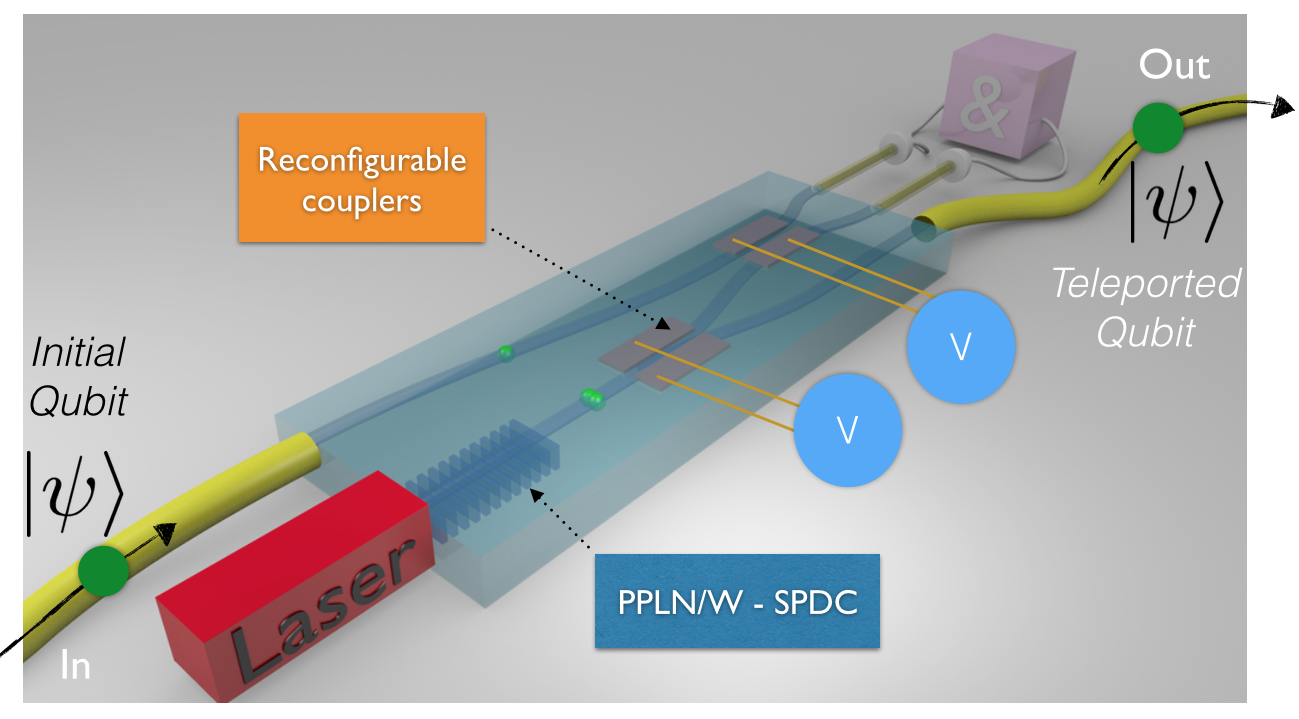}
\caption{On-chip teleportation. The on-chip teleporter developed in Nice in 2012, as a telecom-compatible elementary plug-in. A source of entangled photon pairs and routing circuitry are integrated on a single photonic substrate enabling on-chip teleportation of an incoming photon.}
\end{figure}

As one can see from these insights into current research, many fundamental problems related to entanglement and teleportation remain to be solved. Furthermore, on the application side, the connection of distant nodes into quantum networks, which will enable provably secure communication and networked quantum computers, will remain an important challenge for many years to come. One thing is clear: this highly multi-disciplinary field will continue to be exciting!

\newpage

\section*{References}
{\small
\vspace{-.25cm}
\noindent
[1] N. Gisin, ``Quantum Chance: Nonlocality, Teleportation and Other Quantum Marvels'' (Springer, 2014, ISBN: 978-3319054724).\vspace{0.15cm}

\noindent
[2] A. Aspect, P. Grangier, and G. Roger, ``Experimental Realization of EPR-Bohm Gedankenexperiment: A New Violation of Bell's Inequalities'', Phys. Rev. Lett. \textbf{49}, 91 (1982).\vspace{0.15cm}

\noindent
[3] M. A. Nielsen, and I. L. Chuang, ``Quantum Computation and Quantum Information'' (Cambridge University Press, 2000, ISBN: 978-0521635035).\vspace{0.15cm}

\noindent
[4] W. Tittel, J. Brendel, H. Zbinden, and N. Gisin, ``Violation of Bell Inequalities by Photons More Than 10\,km Apart'', Phys. Rev. Lett. \textbf{81}, 3563 (1998).\vspace{0.15cm}

\noindent
[5] V. Scarani, ``The device-independent outlook on quantum physics'', Acta Phys. Slov. \textbf{62}, 347 (2013); N. Brunner, D. Cavalcanti, S. Pironio, V. Scarani, and S. Wehner, ``Bell nonlocality'', Rev. Mod. Phys. \textbf{86}, 419 (2014).\vspace{0.15cm}

\noindent
[6] H. J. Kimble, ``The Quantum Internet'', Nature \textbf{453}, 1023 (2008).\vspace{0.15cm}

\noindent
[7] J.-D. Bancal, S. Pironio, A. Acin, Y.-C. Liang, V. Scarani, and N. Gisin, ``Quantum non-locality based on finite-speed causal influences leads to superluminal signalling'', Nat. Phys. \textbf{8}, 867 (2012).\vspace{0.15cm}

\noindent
[8] S. Pirandola, J.Eisert, C. Weedbrook, A. Furusawa, and S. L. Braunstein, ``Advances in Quantum Teleportation'', e-print arXiv:1505.07831 (2015).\vspace{0.15cm}

\noindent
[9] I. Marcikic, H. De Riedmatten, W. Tittel, H. Zbinden, and N. Gisin, ``Long-distance teleportation of qubits at telecommunication wavelengths'', Nature \textbf{421}, 509 (2003).\vspace{0.15cm}

\noindent
[10] O. Landry, J. van Houwelingen, A. Beveratos, H. Zbinden, and N. Gisin, ``Quantum teleportation over the Swisscom telecom network'', J. Opt. Soc. Am. B 24 (2), \textbf{398} (2007).\vspace{0.15cm}

\noindent
[11] A. Rubenok J. A. Slater, P. Chan, I. Lucio-Martinez, and W. Tittel, ``Real-World Two-Photon Interference and Proof-of-Principle Quantum Key Distribution Immune to Detector Attacks'', Phys. Rev. Lett. \textbf{111}, 130501 (2013).\vspace{0.15cm}

\noindent
[12] F. Bussi\`eres, C. Clausen, A. Tiranov, B. Korzh, V. B. Verma, S. W. Nam, F. Marsili, A. Ferrier, P. Goldner, H. Herrmann, C. Silberhorn, W. Sohler, M. Afzelius, and N. Gisin, ``Quantum teleportation from a telecom-wavelength photon to a solid-state quantum memory'', Nat. Phot. \textbf{8}, 775 (2014).\vspace{0.15cm}

\noindent
[13] These two demonstrations were published back-to-back: C. Clausen, I. Usmani, F. Bussières, N. Sangouard, M. Afzelius, H. De Riedmatten, and N. Gisin, ``Quantum storage of photonic entanglement in a crystal'', Nature \textbf{469}, 508 (2011), and E. Saglamyurek, N. Sinclair, J. Jin, J. A. Slater, D. Oblak, F. Bussières, M. George, R. Ricken, W. Sohler, and W. Tittel, ``Broadband waveguide quantum memory for entangled photons'', Nature \textbf{469}, 512 (2011).\vspace{0.15cm}

\noindent
[14] X.-L. Wang, X.-D. Cai, Z.-E. Su, M.-C. Chen, D. Wu, L. Li, N.-L. Liu, C.-Y. Lu, and J.-W. Pan, ``Quantum teleportation of multiple degrees of freedom of a single photon'', Nature \textbf{518}, 516 (2015).\vspace{0.15cm}

\noindent
[15] A. Martin, O. Alibart, M. P. De Micheli, D. B. Ostrowsky, and S. Tanzilli, ``A quantum relay chip based on telecommunication integrated optics technology'', New J. Phys. \textbf{14}, 025002 (2012).
}

\newpage
%
%

\end{document}